\documentclass[aps,prd,onecolumn,groupedaddress,showpacs,nofootinbib,amssymb,eqsecnum]{revtex4}
\usepackage{amsmath}
\usepackage{amssymb}
\usepackage{amsfonts}
\usepackage{graphicx,bm}
\usepackage{dcolumn}
\usepackage{epsfig}
\usepackage{color,amsxtra}
\usepackage{epsf}
\usepackage{enumerate}
\usepackage{hhline}
\usepackage{array}
\usepackage{tabularx}
\newcommand{\be}{\begin{equation}}
\newcommand{\ee}{\end{equation}}
\newcommand{\bea}{\begin{eqnarray}}
\newcommand{\eea}{\end{eqnarray}}
\newcommand{\beaa}{\begin{eqnarray*}}
\newcommand{\eeaa}{\end{eqnarray*}}

\newcommand{\nn}{\nonumber \\}
\newcommand{\e}{\mathrm{e}}

\allowdisplaybreaks[4]

\newcommand{\Eqn}[1]{&\hspace{-0.2em}#1\hspace{-0.2em}&}

\def\be{\begin{equation}}
\def\ee{\end{equation}}
\def\bea{\begin{eqnarray}}
\def\eea{\end{eqnarray}}

\def\nn{\nonumber \\}
\def\e{\mathrm{e}}

\begin{document}


\title{Possible antigravity regions in $F(R)$ theory?}
\author{Kazuharu Bamba$^{1,}$\footnote{
E-mail address: bamba@kmi.nagoya-u.ac.jp},
Shin'ichi Nojiri$^{1, 2,}$\footnote{E-mail address:
nojiri@phys.nagoya-u.ac.jp},
Sergei D. Odintsov$^{2, 3, 4, 5}$\footnote{
E-mail address: odintsov@ieec.uab.es}
and
Diego S\'{a}ez-G\'{o}mez$^{6, 7,}$\footnote{
E-mail address: diego.saezgomez@uct.ac.za}
}
\affiliation{
$^1$Kobayashi-Maskawa Institute for the Origin of Particles and the
Universe,
Nagoya University, Nagoya 464-8602, Japan\\
$^2$Department of Physics, Nagoya University, Nagoya 464-8602, Japan\\
$^3$Instituci\`{o} Catalana de Recerca i Estudis Avan\c{c}ats (ICREA),
Barcelona, Spain\\
$^4$Institut de Ciencies de l'Espai (CSIC-IEEC),
Campus UAB, Facultat de Ciencies, Torre C5-Par-2a pl, E-08193 Bellaterra
(Barcelona), Spain\\
$^5$ Tomsk State Pedagogical University, Kievskaya Avenue, 60,
634061, Tomsk, Russia\\
$^6$Astrophysics, Cosmology and Gravity Centre (ACGC) and \\
Department of Mathematics and Applied Mathematics, University of Cape Town, 
Rondebosch 7701, Cape Town, South Africa \\
$^7$Fisika Teorikoaren eta Zientziaren Historia Saila, Zientzia eta Teknologia 
Fakultatea,\\
Euskal Herriko Unibertsitatea, 644 Posta Kutxatila, 48080 Bilbao, Spain
}


\begin{abstract}
We construct an $F(R)$ gravity theory corresponding to the Weyl invariant 
two scalar field theory.
We investigate whether such $F(R)$ gravity can have the
antigravity regions where the Weyl curvature invariant
does not diverge at the Big Bang and Big Crunch singularities. It is 
revealed that the divergence cannot
be evaded completely but can be much milder than that in the original 
Weyl invariant two scalar field theory.
\end{abstract}


\maketitle

\section{Introduction}

Recent observations~\cite{LSS, Eisenstein:2005su, WMAP-PLANCK, Jain:2003tba} 
including Type Ia Supernovae~\cite{SN1}
have suggested the current cosmic expansion
is accelerating. For the universe to be strictly homogeneous and isotropic, 
there are two major approaches: To introduce dark energy within general 
relativity (for reviews, see \cite{DE-R}) and to modify gravity on large 
distances (for recent reviews, see \cite{F(R)-R-NO-CF-CD-DS})
Furthermore, it was realized that modified gravity can describe dark 
energy~\cite{Capozziello:2002rd, Carroll:2003wy} and also unify dark energy era with 
early-time cosmic acceleration \cite{Nojiri:2003ft}.

Theoretical features of such modified gravity theories themselves
become important concerns in the literature.
For instance, the scale invariance in inflationary cosmology~\cite{Kallosh:2013oma, CIIC} or cyclic cosmologies with the Weyl invariant scalar 
fields~\cite{Lubbe:2013hfa}\footnote{For the early universe cosmology 
in the case of two scalar fields not the Weyl invariantly coupled to 
the scalar curvature, see~\cite{Bamba:2006mh}.}
have recently been studied. 
On the other hand, the cosmological transition from gravity to antigravity
has been examined in various background space-time including the strictly
homogeneous and isotropic Friedmann-Lema\^{i}tre-Robertson-Walker (FLRW)
universe~\cite{Linde:1979kf, Starobinsky:1981**, Futamase:1989hb, 
Abramo:2002rn, Caputa:2013mfa}\footnote{We remark that generally speaking the 
antigravity regime is possible in $F(R)$ when its first derivative 
is negative. 
Of course, it leads to number of unpleasant consequences like the possibility 
of only static universe due to the change of 
gravitational coupling constant sign in the FLRW equations. Hence, 
such possibility seems to be rather speculative one which may occur somewhere 
before the Big Bang. 
One can speculate that the Big Bang itself is the transition point from 
antigravity to gravity regimes as due to passing through zero of 
gravitational coupling 
constant some singularity may be expected. 
Also, note that even currently some variation of gravitational 
coupling constant may be expected as discussed 
in recent Ref.~\cite{Bronnikov:2013jua}.}. 
Moreover, in Refs.~\cite{Bars:2011th, Bars:2011aa, B-BST-BST, Bars:2013qna}, 
it has been explored that in extended theories of general relativity with the 
Weyl invariance (or conformal invariance), antigravity regimes 
have to be included. Very recently, it has been verified in 
Ref.~\cite{Carrasco:2013hua} that the Weyl invariant
 becomes infinite at both the Big Bang (Big Crunch)
singularity appearing at the transition from antigravity (gravity) and gravity 
(antigravity).

In this Letter, we reconstruct an $F(R)$ gravity theory corresponding to the 
Weyl invariant two scalar field theory.
Our original motivation is to demonstrate that the Weyl invariant two 
scalar field theory can be reformulated in terms
of $F(R)$ gravity (see, for instance, Ref.~\cite{Elizalde:2008yf}). 
In addition, we examine whether the $F(R)$ gravity can pass through the 
antigravity regions.
We use units of $k_\mathrm{B} = c = \hbar = 1$,
where $c$ is the speed of light, and denote the
gravitational constant $8 \pi G_{\mathrm{N}}$ by
${\kappa}^2 \equiv 8\pi/{M_{\mathrm{Pl}}}^2$
with the Planck mass of $M_{\mathrm{Pl}} = G_{\mathrm{N}}^{-1/2} = 1.2 \times
10^{19}$\,\,GeV.
We also adopt the metric signature $\mathrm{diag} (-, +, +, +)$.

The Letter is organized as follows.
In Sec.\ II, we introduce the Weyl invariant scalar theory and
present it as the corresponding $F(R)$ gravity theory.
In Sec.\ III, we explore how the corresponding $F(R)$ gravity theory obtained 
above can be connected with antigravity regions.
In Sec.\ IV, some conclusions are presented.

\section{The Weyl transformation in $F(R)$ gravity}

\subsection{The Weyl invariant scalar field theory}

An action for the Weyl invariant scalar field theory
in the presence of matter is given by~\cite{B-D}
\begin{equation}
S = \int d^4 x \sqrt{-g}
\left( - \omega f(\phi) R
-\frac{\omega}{2} g^{\mu\nu}
\nabla_{\mu} \phi \nabla_{\nu} \phi
-V(\phi) \right)
+\int d^4 x
{\mathcal{L}}_{\mathrm{M}} \left( g_{\mu\nu}, {\Psi}_{\mathrm{M}} \right)\,,
\label{eq:2.1}
\end{equation}
with
\begin{eqnarray}
f(\phi) \Eqn{=} \frac{1}{2
}
\xi \phi^2 \,, \quad \xi = \frac{1}{6} \,, 
\label{eq:2.2} \\
\omega \Eqn{=} \pm 1 \,,
\label{eq:2.3} \\
V(\phi) \Eqn{=} \frac{\lambda}{4} \phi^4 \,, \quad \lambda =1 \,.
\label{eq:2.4}
\end{eqnarray}
Here, $R$ is the scalar curvature, $g$ is the determinant of the metric 
$g_{\mu\nu}$,
$\nabla_\mu$ is the covariant derivative operator associated
with $g_{\mu\nu}$
(for its operation on a scalar field,
$\nabla_{\mu} \phi = \partial_{\mu} \phi$),
$f(\phi)$ is a non-minimal gravitational coupling
term of $\phi$,
$\omega = +1 (-1)$ is the coefficient of kinetic term of
the canonical (non-canonical scalar) field $\phi$,
$\xi$ is a constant determining whether the theory respects the
Weyl invariance, $V(\phi)$ is the potential for
a scalar field $\phi$, and $\lambda$ is a constant.
$\xi$ 
is
dimensionless and $\phi$ 
has
the [mass] dimension.
Moreover, ${\mathcal{L}}_{\mathrm{M}}$ is the matter Lagrangian,
where ${\Psi}_{\mathrm{M}}$ denotes
all the matter fields such as those in the standard model of particle physics
(and it does not include the scalar field $\phi$).

It is significant to remark that since
the scalar curvature $R$ is represented as
$R = -\left( T + 2\nabla^{\mu} T^{\rho}_{\verb| |\mu \rho}
\right)$~\cite{A-P},
where $T^{\rho}_{\verb| |\mu \rho}$ is the torsion tensor and
$T$ is the torsion scalar in telleparalelism~\cite{Teleparallelism, A-P},
$F(R)$ gravity
is considered to be equivalent to
$F(T+2\nabla^{\mu} T^{\rho}_{\verb| |\mu \rho})$,
and that
the Weyl invariant scalar field theory coupling to
the scalar curvature
is also equivalent to that with its coupling to
the torsion scalar~\cite{Bamba:2013jqa}.

\subsection{The Weyl transformation}

If the Weyl transformation in terms of the action
in Eq.~(\ref{eq:2.1}) is made
as $g_{\mu\nu} \to \hat{g}_{\mu\nu} = \Omega^2 g_{\mu\nu}$,
where $\Omega \equiv \sqrt{f(\phi)}$,
the action in the so-called Jordan frame can be
transformed into that in the Einstein frame~\cite{CT-F-M, Faraoni:2006fx}.
Here, the hat denotes quantities in 
the Einstein frame for the present case.
On the other hand, it is known that
a non-minimal scalar field theory corresponding to an $F(R)$ gravity theory
is the Brans-Dicke theory~\cite{B-D-PRD}
which has the potential term and does not the kinetic term, i.e.,
the Brans-Dicke parameter $\omega_{\mathrm{BD}} =0$.


We examine an $F(R)$ gravity theory corresponding to the Weyl invariant 
scalar field theory.
We now consider the following action given by Eq.~(\ref{eq:2.1})
with $\omega = -1$ and without the matter part
\be
\label{I1a}
S = \int d^4 x \sqrt{-g}
\left[
\frac{\phi^2}{12} R
+\frac{1}{2} g^{\mu\nu} \partial_{\mu} \phi \partial_{\nu} \phi - 
\frac{\lambda}{4} \phi^4 \right] \, .
\ee
First looking this action, one may think the field $\phi$ is ghost since the 
kinetic term is not canonical.
We can, however, remove the ghost because the action (\ref{I1a}) is invariant 
under the Weyl transformation.
By using the Weyl transformation, we may fix the scalar field $\phi$ to be a 
constant,
\be
\label{I1b}
\phi^2 = \frac{6}{\kappa^2}\, .
\ee
Then we obtain the action of the Einstein gravity with cosmological constant:
\be
\label{I1c}
S = \int d^4 x \sqrt{-g}
\left[
\frac{1}{2\kappa^2} R - \frac{9 \lambda}{\kappa^4} \right] \, .
\ee
The action (\ref{I1c}) can be also reproduced 
by using the scale transformation 
$g_{\mu\nu}= \left( \frac{6}{\kappa^2 \phi^2} \right) \hat g_{\mu\nu}$.
In this case,
the scalar curvature is transformed as
\be
\label{I2a}
R = \frac{\phi^2}{6\kappa^2} \left( \hat R + \frac{6\hat\Box \phi}{\phi}
- \frac{12 {\hat g}^{\mu\nu} \partial_\mu \phi \partial_\nu \phi}{\phi^2} 
\right) \, ,
\ee
and hence the action (\ref{I1a}) is represented as
\begin{align}
\label{I3a}
S =& \frac{1}{2\kappa^2}\int d^4 x \sqrt{-\hat g}
\left[ \hat R+ \frac{6\hat\Box \phi}{\phi}
 - \frac{12 {\hat g}^{\mu\nu} \partial_\mu \phi \partial_\nu \phi}{\phi^2}
+\frac{6}{\phi^2} {\hat g}^{\mu\nu} \partial_{\mu} \phi \partial_{\nu} \phi
 - \frac{18 \lambda}{\kappa^2} \right] \nn
=& \int d^4 x \sqrt{- \hat g}
\left[
\frac{1}{2\kappa^2} \hat R - \frac{9 \lambda}{\kappa^4} \right] \, .
\end{align}
Thus, the corresponding $F(R)$ gravity theory is
$F(R) = \left(R - 2\Lambda \right)/\left( 2\kappa^2 \right)$
with $\Lambda \equiv 9 \lambda/\kappa^2$ as in Eq.~(\ref{I3a}),
that is, the Einstein-Hilbert action including cosmological constant.


We mention that
it is meaningful to explore the reason why
the corresponding $F(R)$ gravity theory in Eq.~(\ref{I3a}),
into which the Weyl invariant scalar field
theory is transformed, has no Weyl invariance.
This is because that when we write $\hat g_{\mu\nu}$ as
$\hat g_{\mu\nu} = \left( \frac{\kappa^2 \phi^2}{6} \right) g_{\mu\nu}$,
the theory is trivially invariant under the Weyl transformation: $\phi \to 
\Omega^{-1} \phi$
and $g_{\mu\nu} \to \Omega^2 g_{\mu\nu}$.


\section{Connection with antigravity}

\subsection{The Weyl invariantly coupled two scalar field theory}

We investigate the Weyl invariantly coupled two scalar field theory. 
This was first proposed in Ref.~\cite{Bars:2008sz} 
and cosmology in it was explored in Ref.~\cite{Bars:2010zh}. 
Recently, the connection with the region of antigravity has also been 
examined in Ref.~\cite{Carrasco:2013hua}. 
The action is described
as~\cite{Bars:2008sz, Bars:2010zh, Carrasco:2013hua, Bars:2011th, Bars:2011aa, B-BST-BST, Nishino:2004kb}\footnote{Note that the action of such a sort assuming phantom-like kinetic term for $u$ may be obtained from more general non-conformal theory due to the asymptotical conformal invariance~\cite{B-O, Buchbinder:1992rb}. 
This phenomenon often occurs in asymptotically-free theories.}
\begin{eqnarray}
S \Eqn{=} \int d^4 x \sqrt{-g}
\left[
\frac{\left( \phi^2 - u^2 \right)}{12} R
+\frac{1}{2} g^{\mu\nu} \left( \partial_{\mu} \phi \partial_{\nu} \phi
- \partial_{\mu} u \partial_{\nu} u \right) -\phi^4 J (u/\phi) \right]
\nonumber \\
&&
{}+ \int d^4 x
{\mathcal{L}}_{\mathrm{M}} \left( g_{\mu\nu}, {\Psi}_{\mathrm{M}}
\right)\,,
\label{eq:3.1}
\end{eqnarray}
where $u$ is another scalar field and $J$ is a function of
a quantity $u/\phi$.
The important point is that this action respects the Weyl symmetry,
even though the coefficient of the kinetic term for
the scalar field $\phi$ is a wrong sign, namely, $\phi$ is
not the canonical scalar field in this action.
Indeed, this action is invariant under the Weyl transformations
$\phi \to \Omega \phi$, $u \to \Omega u$, and $g_{\mu\nu} \to
\Omega^{-2} g_{\mu\nu}$.
This implies that there does not exist any ghost.
%
%

\subsection{Representation as single scalar field theory with 
its Weyl invariant coupling}

We rewrite the action in Eq.~(\ref{eq:3.1}) with two scalar fields
to the one described by single scalar field through
the Weyl transformation.

For the action (\ref{eq:3.1}) without the matter part, given by
\begin{equation}
S = \int d^4 x \sqrt{-g}
\left[
\frac{\left( \phi^2 - u^2 \right)}{12} R
+\frac{1}{2} g^{\mu\nu} \left( \partial_{\mu} \phi \partial_{\nu} \phi
- \partial_{\mu} u \partial_{\nu} u \right) -\phi^4 J (u/\phi) \right]\,,
\label{I1}
\end{equation}
we may consider the Weyl transformation
$g_{\mu\nu}= \phi^{-2} \hat g_{\mu\nu}$.
The scalar curvature is transformed as
\be
\label{I2}
R = \phi^2 \left( \hat R + \frac{6\hat\Box \phi}{\phi} - \frac{12 {\hat 
g}^{\mu\nu} \partial_\mu \phi \partial_\nu \phi}
{\phi^2} \right) \, .
\ee
Accordingly, the action (\ref{I1}) is reduced to
\begin{align}
\label{I3}
S =& \int d^4 x \sqrt{-\hat g}
\left[
\frac{1}{12}\left( 1 - \frac{u^2}{\phi^2} \right) \hat R
+ \left( 1 - \frac{u^2}{\phi^2} \right) \left( \frac{\hat\Box \phi}{2\phi}
 - \frac{ {\hat g}^{\mu\nu} \partial_\mu \phi \partial_\nu \phi}{\phi^2} 
\right)
+\frac{1}{2\phi^2} {\hat g}^{\mu\nu} \left( \partial_{\mu} \phi \partial_{\nu} 
\phi
 - \partial_{\mu} u \partial_{\nu} u \right) - J (u/\phi) \right] \nn
=& \int d^4 x \sqrt{-\hat g}
\left[
\frac{1}{12}\left( 1 - \frac{u^2}{\phi^2} \right) \hat R
 - \frac{1}{2} {\hat g}^{\mu\nu} \partial_{\mu} \left(\frac{u}{\phi}\right) 
\partial_{\nu} \left(\frac{u}{\phi}\right)
 - J (u/\phi) \right] \, .
\end{align}
Therefore if we define a new scalar field $\varphi \equiv u/\phi$, the action 
has the following form:
\be
\label{I4}
S = \int d^4 x \sqrt{-\hat g}
\left[
\frac{1}{12}\left( 1 - \varphi^2 \right) \hat R
 - \frac{1}{2} {\hat g}^{\mu\nu} \partial_{\mu} \varphi \partial_{\nu} \varphi
 - J (\varphi) \right] \, .
\ee
The obtained action has no Weyl invariance because ${\hat g}_{\mu\nu}$ 
and $\varphi$ are invariant
under the Weyl transformation. The Weyl invariance appears because we write
${\hat g}_{\mu\nu}= \phi^2 g_{\mu\nu}$ and $\varphi = u/\phi$. Therefore the 
Weyl invariance is artificial
or fake, or hidden local symmetry.
Conversely even in an arbitrary $F(R)$ gravity, 
if we write the metric as $g_{\mu\nu}= 
\phi^2 {\tilde g}_{\mu\nu}$,
there always appears the Weyl invariance.

\subsection{Corresponding $F(R)$ gravity}

We may relate the action (\ref{I4}) with $F(R)$ gravity.
By the further Weyl transforming the metric as ${\hat 
g}_{\mu\nu}=\e^{\eta(\varphi)} {\bar g}_{\mu\nu}$ with
a function $\eta(\varphi)$, we rewrite the action (\ref{I4}) in the following 
form:
\be
\label{I5}
S = \int d^4 x \sqrt{-\bar g}
\left[
\frac{\e^{\eta(\varphi)}}{12}\left( 1 - \varphi^2 \right) \bar R
 - \frac{\e^{\eta(\varphi)}}{8} {\bar g}^{\mu\nu} \left( \left( 1 + 2 \varphi^2 
\right) \eta' (\varphi) ^2
 - 4 \eta' (\varphi) - 4 \right) \partial_{\mu} \varphi \partial_{\nu} \varphi
 - \e^{2\eta(\varphi)}J (\varphi) \right] \, ,
\ee
where the prime means the derivative with respect to $\varphi$, 
and the bar shows the quantities 
after the above Weyl 
transformation.
Then if choose $\eta(\varphi)$ by
\be
\label{I6}
\left( 1 + 2 \varphi^2 \right) \eta' (\varphi) ^2 - 4 \eta' (\varphi) - 4 = 0\, 
, \quad \mbox{that is} \quad
\eta'(\varphi) = \frac{2\varphi \pm 2 \sqrt{ 3\varphi^2 + 1}}{ 1 + 2 
\varphi^2}\, ,
\ee
the kinetic term of $\varphi$ vanishes and we obtain
\be
\label{I7}
S = \int d^4 x \sqrt{-\bar g}
\left[
\frac{\e^{\eta(\varphi)}}{12}\left( 1 - \varphi^2 \right) \bar R
 - \e^{2\eta(\varphi)}J (\varphi) \right] \, .
\ee
Then by the variation of the action with respect to $\varphi$, we obtain an 
algebraic equation,
which can be solved with respect to $\varphi$ as a function of $\bar{R}$, 
$\varphi = \varphi(\bar{R})$.
Then by substituting the expression into the action (\ref{I7}), we obtain an 
$F(R)$ gravity:
\be
S = \int d^4 x \sqrt{-\bar g} F(\bar{R})\, ,\quad
F(\bar{R}) = \frac{\e^{\eta\left(\varphi\left(\bar{R}\right)\right)}}{12}\left( 
1 - \varphi\left(\bar{R}\right)^2 \right) \bar R
 - \e^{2\eta\left(\varphi\left(\bar{R}\right)\right)}J 
\left(\varphi\left(\bar{R}\right)\right) \, .
\label{eq:I8}
\ee

\subsection{Finite-time future singularities}

Now let us examine the reconstruction of the above model when a singular flat FLRW cosmology is considered. In this background, the metric is given by $ds^2=-dt^2+a(t)^2\sum_{i=1}^3 \left(dx^{i}\right)^2$, where $a(t)$ the scale factor. Depending on the nature of the singularity, a classification of finite-time future singularities in the FLRW cosmologies was presented in Ref.~\cite{Nojiri:2005sx} as follows. 
\begin{itemize}
\item Type I (``Big Rip''): For $t\rightarrow t_{\mathrm{s}}$, $a\rightarrow
\infty$ and $\rho\rightarrow \infty$, $|P|\rightarrow \infty$.
\item Type II (``Sudden''): For $t\rightarrow t_{\mathrm{s}}$, $a\rightarrow
a_s$ and $\rho\rightarrow \rho_{\mathrm{s}}$, $|P|\rightarrow \infty$.
\item Type III: For $t\rightarrow t_{\mathrm{s}}$, $a\rightarrow a_{\mathrm{s}}$
and $\rho\rightarrow \infty$, $|P|\rightarrow \infty$.
\item Type IV: For $t\rightarrow t_{\mathrm{s}}$, $a\rightarrow a_{\mathrm{s}}$ and 
$\rho\rightarrow \rho_{\mathrm{s}}$, $P \rightarrow P_{\mathrm{s}}$ 
but higher derivatives of the Hubble parameter diverge. 
\end{itemize}
Here, $\rho$ and $P$ are the energy density and pressure of the universe, 
respectively. 
We might now study a simple case, where the Hubble parameter 
$H \equiv \dot{a}/a$ is described by 
\be
H=\frac{\alpha}{t_{\mathrm{s}}-t}\ .
\label{D1}
\ee
This solution describes a Big Rip singularity that occurs in a time $t_{\mathrm{s}}$. Then, by the $F(R)$ FLRW equations, the corresponding action (3.9) with the matter action can be reconstructed as 
\bea
H^2 \Eqn{=} \frac{1}{3F_R}\left[\kappa^2 \rho_{\mathrm{M}} +\frac{RF_R-F}{2}-3H\dot{R}F_{RR}\right]\ , \nn
-3H^2-2\dot{H} \Eqn{=} \frac{1}{F_R}\left[\kappa^2 p_{\mathrm{M}} +\dot{R}^2F_{RRR}+2H\dot{R}F_{RR}+\ddot{R}F_{RR}+\frac{1}{2}(F-RF_R)\right]\ ,
\label{D2}
\eea
where the subscripts correspond to derivatives with respect to $R$, and 
$\rho_{\mathrm{M}}$ and $p_{\mathrm{M}}$ are the energy density and 
pressure of all the matters, respectively. 

For the solution (\ref{D1}), it is straightforward to check that the $F(R)$ function yields
\be
F(R)=R^n\ , \quad \text{where} \quad \frac{1-3n+2n^2}{n-2}=\alpha\ , 
\label{D3}
\ee
with $n$ a constant 
Then, by (3.9) the corresponding scalar-tensor theory is obtained as 
\bea
\frac{\e^{\eta(\varphi)}}{12}\left(1-\varphi^2\right) \Eqn{=} \frac{\partial F}{\partial R}=nR^{n-1}\nn
\e^{2\eta(\varphi)}J(\varphi) \Eqn{=} \frac{\partial F}{\partial R}R-F(R)=\left(n-1\right)R^n\ .
\label{D4}
\eea
Thus, the cosmological evolution for the scalar field $\varphi$ is obtained as well as its self-interacting term $J(\varphi)$, such that the corresponding action is obtained. Note that in such a case, the antigravity regime is never crossed, since $R>0$ for (\ref{D1}) which leads to $|\varphi|<1$. Nevertheless, for other kind of singular solutions within the FLRW metrics, the antigravity regime might be expected.

\subsection{Connection with antigravity}

It is clear from the action of a scalar field theory in Eq.~(\ref{I4}) that if 
$\varphi^2>1$, there emerges antigravity. When this condition is satisfied,
it follows from the form of the corresponding $F(R)$ gravity in 
Eq.~(\ref{eq:I8}) that the coefficient of $\bar{R}$ can be positive as
$\left[ \e^{\eta\left(\varphi\left(\bar{R}\right)\right)}/12 \right] 
\left( 1 - 
\varphi\left(\bar{R}\right)^2 \right) 
< 0$,
and thus antigravity can appear.
In other words,
the effective Newton 
coupling in the action of the corresponding $F(R)$ gravity theory in 
Eq.~(\ref{eq:I8}) is described as
$\bar{G}_{\mathrm{N}}
\equiv 6 \e^{-\eta(\varphi)} G_{\mathrm{N}}/ \left(1- \varphi^2\right)$.
Accordingly, when $\varphi = -1$ and $\varphi = +1$, there happens transitions
between gravity and antigravity. 

We investigate what happens in the travel to the antigravity region for
the $F(R)$ gravity theory in Eq.~(\ref{eq:I8}) corresponding to the Weyl 
invariantly coupled two scalar field theory in Eq.~(\ref{I1}).
By following the procedures in Ref.~\cite{Kallosh:2013oma},
we explore the behaviors of solutions in the anisotropic background metric
so that homogeneous and isotropic solutions should not be singular around
the boundary between gravity and antigravity regions
(for the detailed analysis on homogeneous and isotropic solutions in 
non-minimally coupled scalar field theories, see, e.g.,~\cite{Bars:2012mt}).

Provided that the background metric of the space-time is expressed 
as~\cite{Misner:1969**}
$ds^2 = a^2 (\tau) \left( -d \tau^2 + \sum_{i=1}^{3} \e^{\beta_i} dx_i^2
\right)$ with $\beta_1 \equiv \sqrt{2/3} \alpha_1 (\tau) + \sqrt{2} \alpha_2 
(\tau)$, $\beta_2 \equiv \sqrt{2/3} \alpha_1 (\tau) - \sqrt{2} \alpha_2 
(\tau)$,
and $\beta_3 \equiv -2\sqrt{2/3} \alpha_1 (\tau)$,
where 
$\tau$ is the conformal time, and an anisotropy 
function $\alpha_i$ $(i=1, \cdots, 3)$ only depends on $\tau$.
In the following, the so-called $\gamma$ gauge of
$-g \equiv - \det g_{\mu\nu} = 1$~\cite{Bars:2011th}.
Moreover, we take into account the existence of radiation.
In this gauge, $\phi$ and $u$ are given by~\cite{Bars:2011aa}
\begin{eqnarray}
\phi \Eqn{=} \left(
\frac{\sqrt{\mathcal{C}}}{|\mathcal{C}|^q} \left|\frac{\tau}{\sqrt{6}}\right|^q 
\mathcal{A} |\mathcal{A}|^{-q} \right)
+\left(\frac{2|\mathcal{C}|^q}{\sqrt{\mathcal{C}}} \frac{\tau}{\sqrt{6}}
\left|\frac{\tau}{\sqrt{6}}\right|^{-q} |\mathcal{A}|^{q} \right)\,,
\label{eq:ADD-FR9-8-IIID-01} \\
u \Eqn{=} \left(
\frac{\sqrt{\mathcal{C}}}{|\mathcal{C}|^q} \left|\frac{\tau}{\sqrt{6}}\right|^q 
\mathcal{A} |\mathcal{A}|^{-q} \right)
-\left(\frac{2|\mathcal{C}|^q}{\sqrt{\mathcal{C}}} \frac{\tau}{\sqrt{6}}
\left|\frac{\tau}{\sqrt{6}}\right|^{-q} |\mathcal{A}|^{q} \right)\,,
\label{eq:ADD-FR9-8-IIID-02}
\end{eqnarray}
with
\begin{eqnarray}
\mathcal{A} \Eqn{\equiv}
p+\frac{\rho_{\mathrm{r}}}{\sqrt{6}} \tau = \frac{\rho_{\mathrm{r}}}{\sqrt{6}} 
\left(\tau-\tau_{\mathrm{BC}} \right)\,,
\label{eq:ADD-FR9-8-IIID-03} \\
\tau_{\mathrm{BC}} \Eqn{\equiv} -\frac{\sqrt{6} p}{\rho_{\mathrm{r}}} \,,
\label{eq:ADD-FR9-8-IIID-04} \\
q \Eqn{\equiv} \frac{1}{2} \left(1 + \frac{p_\sigma}{\sqrt{p_\sigma^2 + p_1^2 + 
p_2^2}} \right)\,,
\label{eq:ADD-FR9-8-IIID-05} \\
p \Eqn{\equiv}
\sqrt{p_\sigma^2 + p_1^2 + p_2^2}\,,
\label{eq:ADD-FR9-8-IIID-06}
\end{eqnarray}
where $\mathcal{C}$ is a constant,
$\rho_{\mathrm{r}}$ is a constant originating from
the existence of radiation, and ($p_\sigma$, $p_1$, $p_2$) are constants (the 
case $p_1 = p_2 =0$ is not considered, because in that case $\alpha_1$ and 
$\alpha_2$ becomes constants).
In the limit of $\tau \to 0$, namely, the Big Bang singularity, $\tau/\sqrt{6} 
\to 0$,
while in the limit of $\tau \to \tau_{\mathrm{BC}}$, namely, the Big Crunch 
singularity, $\mathcal{A} \to 0$.
Furthermore, from Eq.~(\ref{eq:ADD-FR9-8-IIID-05})
we find $0 \leq q \leq 1$.

For the $F(R)$ gravity theory whose action is given by Eq.~(\ref{eq:I8}),
the Weyl curvature invariant is considered to be
\begin{equation}
\mathcal{I} = \left[ \frac{\e^{\eta}}{6} \left(1-\varphi^2 \right) \right]^{-2}
C_{\mu\nu\rho\sigma} C^{\mu\nu\rho\sigma}\,,
\label{eq:ADD-FR9-8-IIID-07}
\end{equation}
where $C^{\mu}_{\,\,\nu\rho\sigma}$ is the Weyl curvature tensor.
As a consequence, we acquire
$\mathcal{I} = 243 \e^{-2 \eta} \Upsilon \left(\tau/\sqrt{6}\right)^{\delta_1} 
\mathcal{A}^{\delta_2}$ with $\delta_1 <0$ and $\delta_2 <0$,
where $\Upsilon$ is a function of several variables as $\Upsilon = \Upsilon 
(\tau/\sqrt{6}, p, p_1, p_2, \rho_\mathrm{r})$~\cite{Kallosh:2013oma}.
Thus, at the Big Bang singularity we obtain
$\mathcal{I}|_{\tau \to 0} \to \infty$ because of $\delta_1 <0$,
whereas at the Big Crunch singularity we have
$\mathcal{I}|_{\tau \to \tau_{\mathrm{BC}}} \to \infty$ owing to
$\delta_2 <0$.

It is worthy to emphasize that 
for the original Weyl invariantly coupled 
two scalar field theory~\cite{Carrasco:2013hua} whose action is given by
Eq.~(\ref{eq:3.1}), the power of $\left( \tau/\sqrt{6} \right)$ and that of
$\left(\tau-\tau_{\mathrm{BC}} \right)$, to which $\mathcal{A}$ is
proportional, are equal to ``$-6$'',
while for the present $F(R)$ gravity theory, $-6< \delta_1 <0$ and 
$-6 <\delta_2 <0$, that is, how singular $\mathcal{I}$ is can be much milder 
than that in the original two scalar field 
theory with those Weyl invariant couplings. 

In addition, 
it is interesting to mention that in Ref.~\cite{Bars:2013qna}, 
the following counter-discussions to the statements 
of Ref.~\cite{Carrasco:2013hua} have been presented. 
For a geodesically complete universe, 
it is necessary to match the values of 
all the physical quantities including the divergent curvatures 
with continuous geodesics in the two regions, 
not to prevent the divergence of the curvature at the transition point\footnote{Note that in the same spirit, the possibility to continue the universe evolution through the mild finite-time future singularities like Type IV singularity seems to exist as geodesics may be continued through the singularity.}. 
This has been demonstrated 
through the identification of conserved quantities across 
the transition~\cite{Bars:2013qna} 
and it is not specific but generic consequence. 
Adopting this point of view, the transition through antigravity region in the Weyl invariant scalar field theory (as well as in the above $F(R)$ theory 
seems to be possible. 

\section{Conclusions}

In the present Letter, we have performed the reconstruction of an $F(R)$ gravity theory corresponding to the Weyl invariant two scalar field theory.
We have also demonstrated how the $F(R)$ gravity theory cannot connect with
antigravity region in order for the Weyl invariant to be finite
at the Big Bang and Big Crunch singularities. Nevertheless, the Weyl invariant 
divergence at these singularities can be much milder than that in the original 
Weyl invariantly invariant two scalar field theory. 
It would be very interesting to investigate this problem for 
$F(R)$ bigravity theories \cite{NOS-BAANO} where the above phenomena could 
qualitatively be different due to possible exchanges of gravity-antigravity 
regions between $g$ and $f$ $F(R)$ gravities.

Finally, we mention the way for the energy conditions to be met in our model 
by following the discussions in Ref.~\cite{Capozziello:2013vna}, where 
a novel formulation to deal with additional degrees of freedom appearing 
in extended gravity theories has been made. 
In this work, we have examined the Weyl invariant (two) scalar field theories. 
These can be categorized to non-minimal scalar field theories such as 
the Brans-Dicke theory~\cite{B-D-PRD}, into which $F(R)$ gravity theories 
can be transformed via the conformal transformation. 
According to the consequences found in Ref.~\cite{Capozziello:2013vna}, 
the four (i.e., null, dominant, strong, weak) energy conditions 
can be described as in general relativity, although the physical meanings 
become different from those in general relativity. This is because 
the properties of gravity interactions as well as 
the geodesic and causal structures in modified gravity 
would be changed from those in general relativity. 
Thus, these differences are considered to be significant 
when it is examined whether extended theories of gravity can pass 
the solar-system tests and cosmological constraints. 

\section*{Acknowledgments}

S.D.O. acknowledges the Japan Society for the Promotion of Science (JSPS)
Short Term Visitor Program S-13131 
and the very hearty hospitality at Nagoya University,
where the work was developed.
The work is supported in part
by the JSPS Grant-in-Aid for
Young Scientists (B) \# 25800136 (K.B.);\
that for Scientific Research
(S) \# 22224003 and (C) \# 23540296 (S.N.);\
and
MINECO (Spain), FIS2010-15640 and
AGAUR (Generalitat de Ca\-ta\-lu\-nya), contract 2009SGR-345,
and MES project 2.1839.2011 (Russia)
(S.D.O.).
D. S.-G. also acknowledges the support from the University of the Basque
Country, Project Consolider CPAN Bo. CSD2007-00042 (Spain) and the URC
financial support from the University of Cape Town (South Africa). 


\end{document}